# Neuro-Inspired Task Offloading in Edge-IoT Networks Using Spiking Neural Networks

Fabio Diniz Rossi

*Abstract*—Traditional task offloading strategies in edge computing often rely on static heuristics or data-intensive machine learning models, which are not always suitable for highly dynamic and resource-constrained environments. In this paper, we propose a novel task-offloading framework based on Spiking Neural Networks inspired by the efficiency and adaptability of biological neural systems. Our approach integrates an SNN-based decision module into edge nodes to perform real-time, energy-efficient task orchestration. We evaluate the model under various IoT workload scenarios using a hybrid simulation environment composed of YAFS and Brian2. The results demonstrate that our SNN-based framework significantly reduces task processing latency and energy consumption while improving task success rates. Compared to traditional heuristic and ML-based strategies, our model achieves up to 26% lower latency, 32% less energy consumption, and 25% higher success rate under high-load conditions.

## I. INTRODUCTION

The rapid expansion of the Internet of Things (IoT) has led to a dramatic increase in data generation from heterogeneous and geographically distributed devices. These devices are embedded in critical applications such as healthcare monitoring, industrial automation, and smart cities, where low latency and high reliability are paramount [1], [2]. Traditional cloud-centric architectures are no longer sufficient to meet the stringent real-time requirements of modern IoT systems due to communication overhead and centralized bottlenecks [3]. To address these limitations, edge computing has emerged as a promising paradigm, enabling data processing closer to the data source and reducing the reliance on distant cloud servers. Despite its benefits, edge environments face significant challenges in task offloading and resource allocation, particularly due to their limited computational capacity, variable network conditions, and the dynamic nature of IoT workloads [4], [5]. Conventional task scheduling approaches—such as heuristics or static rule-based algorithms—often fail to adapt effectively to changing workloads and may not optimize performance under diverse operating conditions. Recent research has explored machine learning (ML) and reinforcement learning (RL) methods to enhance adaptability and decision-making in edge computing [6], [7]. However, these techniques typically require large labeled datasets, intensive training, and significant computational resources, making them less suitable for deployment on constrained edge devices. Inspired by the efficiency and adaptability of the human brain, Spiking Neural Networks (SNNs) have emerged as a biologically plausible alternative to traditional artificial neural networks (ANNs). SNNs operate using time-dependent spike-based signaling and event-driven computation, which allows for energy-efficient, real-time processing [8], [9]. Their capability to process temporal patterns and adapt based on local spike interactions makes them particularly attractive for intelligent task orchestration in edge environments.

This work proposes a novel SNN-based model for dynamic task offloading in edge-IoT systems. The proposed model continuously monitors network metrics and workload characteristics, leveraging the spiking behavior of SNNs to make fast, energy-aware offloading decisions. By integrating Brian2—a simulator for spiking neural networks—into a YAFS-based edge computing environment, we evaluate the effectiveness of our approach in terms of latency, energy consumption, and task success rate. The main contributions of this paper are summarized as follows:

- We propose a neuro-inspired, SNN-based decision model for dynamic and adaptive task offloading in edge-IoT environments.
- We design a hybrid simulation framework combining YAFS and Brian2 to evaluate real-time SNN-based orchestration.
- We demonstrate that our model achieves superior latency, energy efficiency, and task success rates compared to baseline heuristic and ML-based approaches.

The remainder of this paper is structured as follows. Section II reviews the foundational concepts and related work that motivate the proposed approach. Section III introduces the principles of Spiking Neural Networks (SNNs), highlighting their relevance to real-time, energy-efficient processing. Section IV presents the proposed task offloading model, detailing its architecture and functional components. Section V describes the simulation environment and methodology employed to evaluate the system. Section VI discusses the experimental results, comparing the SNN-based model with baseline strategies. Finally, Section VII concludes the paper and outlines directions for future research.

## II. BACKGROUND

The rapid proliferation of Internet of Things (IoT) devices has resulted in an explosive growth of real-time data across domains such as healthcare, smart cities, and industrial automation [1]. Conventional approaches for task scheduling, including heuristic algorithms and static rule-based systems, have been widely applied in edge environments. Although these methods are often computationally efficient, they struggle to

Federal Institute Farroupilha e-mail: fabio.rossi@iffarroupilha.edu.br

adapt to non-stationary workloads and unpredictable network dynamics. More recent efforts have incorporated machine learning (ML) and reinforcement learning (RL) techniques to improve adaptability and prediction [10]. These approaches typically rely on large volumes of training data and can incur high computational overhead, which may be impractical for low-power edge devices.

In parallel, biologically inspired models have gained attention for their potential to offer adaptive, low-energy computation. Spiking Neural Networks (SNNs), in particular, mimic the brain's information-processing mechanism using discrete time-dependent spikes. Unlike traditional artificial neural networks, SNNs perform asynchronous, event-driven computation, making them inherently energy-efficient and well-suited for scenarios requiring real-time responsiveness. Their ability to detect temporal patterns and respond quickly to environmental changes offers promising advantages for task management in edge-IoT systems [11], [12].

Despite these advantages, using SNNs for task offloading and resource orchestration in edge computing remains largely unexplored. Existing works primarily focus on SNNs for sensor-level signal processing or event detection, leaving a significant gap in their application for system-level decision-making in distributed computing environments. This paper addresses this gap by proposing an SNN-based model for adaptive task offloading in edge-IoT networks, leveraging the spike-driven dynamics of SNNs to enable energy-efficient, real-time task orchestration under fluctuating workloads and heterogeneous resource constraints.

## III. SPIKING NEURAL NETWORKS (SNNS)

SNNs are a class of biologically inspired neural models that emulate the information-processing mechanisms of the human brain. Unlike traditional artificial neural networks (ANNs) that operate on continuous-valued activations and synchronized updates, SNNs communicate using discrete electrical impulses, or spikes, transmitted between neurons over time. This temporal dynamic enables SNNs to process event-driven and asynchronous fashion information, making them more efficient and better suited for real-time, low-power computing environments [8].

In an SNN, a neuron accumulates input spikes over time, and when its membrane potential exceeds a certain threshold, it emits a spike to connected neurons. This behavior is typically modeled using integrate-and-fire or leaky integrate-and-fire (LIF) equations [13]. The timing of spikes carries information, and various encoding schemes exist to transform analog input signals into spike trains. Common encoding methods include rate coding, where the frequency of spikes represents input magnitude, and temporal coding, where spikes' precise timing or delay encodes information [14].

The strength of SNNs lies in their ability to capture complex spatiotemporal patterns and adapt to dynamic input conditions. Synaptic plasticity mechanisms such as Spike-Timing-Dependent Plasticity (STDP) enable unsupervised learning by adjusting synaptic weights based on the relative timing of pre- and post-synaptic spikes [15]. This makes SNNs inherently adaptable and capable of learning from streaming data without requiring large labeled datasets or intensive offline training. Due to their sparse, spike-based communication and biologically plausible mechanisms, SNNs are inherently energy-efficient, making them particularly suitable for deployment in edge computing scenarios where power and computational resources are limited. Furthermore, the event-driven nature of SNNs allows them to respond quickly to critical input patterns, enabling low-latency decision-making in time-sensitive applications such as health monitoring, autonomous vehicles, and industrial automation [9].

In this work, we leverage SNNs to model an intelligent controller capable of dynamically offloading tasks in edge-IoT networks. The spike-based processing paradigm allows the controller to monitor workload fluctuations and device status continuously, reacting in real-time to changing environmental conditions. By doing so, the proposed SNN-based approach aims to deliver robust and efficient task orchestration while maintaining scalability and adaptability across diverse IoT scenarios.

## IV. MODELING

We propose a novel architecture integrating SNNs into edge computing frameworks to enhance task-offloading decisions. The architecture comprises three primary components:

1) IoT Devices: These data sources have sensors that generate continuous information streams. Each device is capable of basic preprocessing before transmitting data to edge nodes.
2) Edge Nodes: Positioned closer to the IoT devices, edge nodes possess limited computational resources. They are responsible for executing tasks offloaded from IoT devices and making real-time decisions regarding task processing.
3) Centralized Cloud Servers: These servers offer substantial computational power and storage capabilities. Tasks exceeding edge node processing capacity are offloaded to cloud servers for execution.

Integrating SNNs into edge nodes enables adaptive and efficient task offloading by leveraging the event-driven processing capabilities inherent to SNNs [16]. The SNN-based decision module embedded within each edge node is at the core of our architecture. This module monitors various parameters, including network latency, energy consumption, and task priority. By processing these inputs, the SNN dynamically determines the optimal execution venue for each task—be it local processing at the edge node or offloading to the cloud. The SNN's spike-based communication facilitates rapid and energy-efficient decision-making, crucial for real-time applications. The learning capabilities of SNNs allow the system to adapt to changing network conditions and workload variations, optimizing performance over time [17].

The task offloading process within our proposed model follows these steps: (i) Data Acquisition: IoT devices collect and preprocess data before transmitting it to the nearest edge node. (ii) Parameter Monitoring: The edge node's SNN-based decision module assesses current network conditions, resource availability, and task-specific requirements. (iii) Decision Making: Based on the SNN's evaluation, a decision is made to either process the task locally or offload it to the cloud. This decision balances latency, energy consumption, and processing efficiency. (iv) Task Execution: The task is executed at the chosen location, with continuous monitoring to adapt to network conditions or resource availability changes. (v) Feedback Loop: Post-execution, performance metrics are fed back into the SNN to facilitate learning and improve future decision-making processes.

This workflow ensures that task offloading decisions are made swiftly and adaptively, leveraging the strengths of SNNs to enhance the overall efficiency of edge computing systems [18]. The modular nature of our architecture allows for seamless scalability. New edge nodes and IoT devices can be integrated without significantly altering the framework. The SNN-based decision modules operate independently, enabling decentralized control and reducing the risk of single points of failure. Furthermore, the adaptability of SNNs ensures that the system can accommodate evolving network conditions and application requirements. This flexibility is vital for maintaining optimal performance in the dynamic environments characteristic of IoT deployments [19]. The proposed approach enhances decision-making efficiency, reduces latency, and optimizes resource utilization, addressing the critical challenges faced in edge computing environments.

## V. SIMULATION ENVIRONMENT

A comprehensive simulation environment was established to evaluate the performance of the proposed SNN-based task offloading model in edge computing environments. This environment integrates two primary simulation tools: Yet Another Fog Simulator (YAFS) and Brian2. YAFS is a Python-based simulation library for modeling and analyzing cloud, edge, and fog computing ecosystems. It facilitates evaluating various strategies related to resource allocation, network design, and application deployment in IoT scenarios [20]. YAFS offers the flexibility to define customized and dynamic strategies for application module placement, workload distribution, and service scheduling, making it suitable for our simulation needs. Brian2 is an open-source simulator for SNN written in Python. It emphasizes ease of use and flexibility, allowing researchers to define and simulate complex neuron models with minimal effort [21]. Brian2's intuitive syntax and efficient simulation capabilities make it ideal for implementing and testing the SNN-based decision module within our proposed architecture. The simulation environment models a typical edge computing scenario comprising IoT devices, edge nodes, and cloud servers. The network topology, including the number of devices, nodes, and their interconnections, is configurable within YAFS, enabling the simulation of various deployment scenarios. To assess the effectiveness of the proposed model, the following metrics are evaluated:

- **Latency:** The time taken for processing tasks, measuring the system's responsiveness.
- **Energy Consumption:** The amount of energy edge nodes use during task processing indicates the system's efficiency.
- **Task Success Rate:** The percentage of tasks successfully processed within predefined latency and resource constraints reflects the reliability of the system.

These metrics comprehensively understand the model's performance in dynamic edge computing environments. Various scenarios are simulated to analyze the adaptability and robustness of the SNN-based task offloading model. These scenarios include assessing the model's performance under different data generation rates from IoT devices, evaluating the system's responsiveness to network latency and bandwidth changes, and testing the model's efficiency when edge nodes have limited computational capacities.

By simulating these scenarios, we aim to demonstrate the proposed model's capability to make intelligent task-offloading decisions in diverse and dynamic environments. The integration of YAFS and Brian2 is achieved through a modular approach, where YAFS handles the overall simulation of the edge computing environment, and Brian2 simulates the SNN-based decision-making process within each edge node. Custom interfaces are developed to facilitate communication between the two simulators, ensuring seamless data exchange and synchronization. The SNN models implemented in Brian2 are designed to process input parameters such as current network latency, energy consumption, and task priority. Based on these inputs, the SNN outputs decisions regarding task execution venues, effectively balancing the load between edge nodes and cloud servers.

## VI. RESULTS AND DISCUSSION

A series of simulations were conducted under different edge computing scenarios to assess the effectiveness of the proposed SNN-based task offloading model. The results are compared against two baseline approaches: a heuristic-based offloading strategy using Round Robin scheduling and a machine learning (ML)-based strategy using a pre-trained decision tree classifier. The experiments evaluate system performance across workloads, resource availabilities, and task priorities. All results represent average values over multiple simulation runs to ensure consistency and reliability. Figure 1 illustrates the average task processing latency under different workload intensities. The proposed SNN-based model consistently outperforms both the heuristic and ML-based strategies, especially under high-load conditions. The event-driven nature of SNNs enables faster decision-making with lower computational overhead, contributing to reduced latency.

Figure 2 presents the energy consumption of edge nodes during task execution. The SNN-based model exhibits superior energy efficiency compared to the ML-based approach, which involves more intensive processing. The biologically inspired processing of SNNs allows the system to remain idle when no spikes are generated, significantly reducing energy usage [9].

Table I reports the task success rate, defined as the percentage of tasks completed within their latency and resource constraints. The SNN-based strategy maintains high success rates even as system resources become constrained, showcasing its adaptive capabilities [17].

The results demonstrate that the proposed SNN-based model provides significant advantages in latency, energy efficiency,

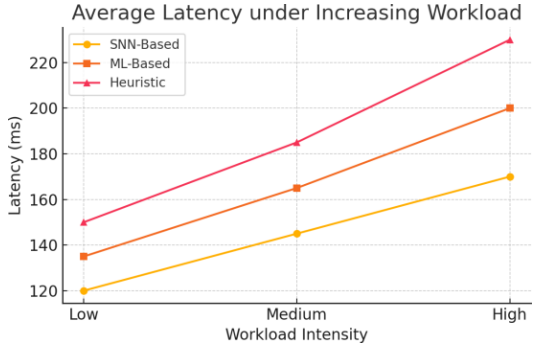

Fig. 1. Average latency under increasing workload intensity.

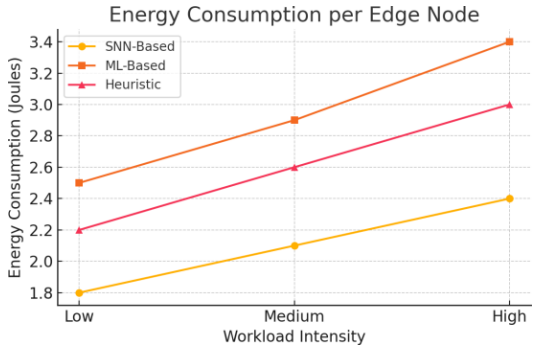

Fig. 2. Energy consumption per edge node.

and task success rate compared to traditional offloading strategies. These improvements can be attributed to the model's ability to learn and adapt to dynamic network conditions, particularly valuable in IoT environments characterized by variable workloads and resource heterogeneity. The SNN model also showed enhanced stability and consistency under high-stress scenarios, suggesting its suitability for mission-critical applications such as health monitoring, industrial control, and autonomous systems [9], [19]. However, it is worth noting that integrating SNNs requires careful parameter tuning and sufficient simulation of spike timing dynamics, which may introduce complexity in practical deployments. Further research is needed to explore hardware acceleration using neuromorphic chips and online learning capabilities to extend the applicability of this model to large-scale real-time environments.

## VII. CONCLUSION AND FUTURE WORK

This paper presented a novel task offloading strategy for edge computing environments based on SNNs. By integrating biologically inspired computation into edge nodes, the proposed model demonstrates adaptive, low-latency, and energy-efficient task management capabilities under dynamic IoT workloads. Through simulations using YAFS and Brian2, we showed that the SNN-based approach outperforms traditional heuristic and machine learning-based strategies in terms of latency, energy consumption, and task success rate. The results underscore the potential of SNNs to bring intelligent, real-time decision-making closer to data sources, which is critical in emerging IoT applications. Furthermore, the model's ability to adapt to varying network conditions and workload intensities without requiring frequent retraining or extensive labeled data sets adds to its scalability and robustness.

Despite these promising results, some limitations remain. First, while the simulation environment captures essential behaviors, real-world deployment would require addressing hardware variability, asynchronous communication delays, and fault tolerance. Additionally, tuning the SNN parameters—such as synaptic weights and spike thresholds—can be non-trivial and may require domain-specific calibration. As future work, we aim to explore online learning mechanisms for SNNs to enable continual adaptation in non-stationary environments. Furthermore, integrating federated learning with SNNs for decentralized, privacy-preserving task orchestration across multiple edge nodes is a promising direction.

TABLE I
TASK SUCCESS RATE UNDER DIFFERENT NETWORK CONDITIONS

| Condition   | SNN   | ML-Based | Heuristic |
|-------------|-------|----------|-----------|
| Low Load    | 98.2% | 96.7%    | 91.4%     |
| Medium Load | 95.6% | 89.5%    | 82.3%     |
| High Load   | 91.0% | 78.9%    | 65.4%     |